\documentclass{SciPost}

\hypersetup{
    colorlinks,
    linkcolor={red!50!black},
    citecolor={blue!50!black},
    urlcolor={blue!80!black}
}

\usepackage[bitstream-charter]{mathdesign}
\usepackage{orcidlink} 

\DeclareSymbolFont{usualmathcal}{OMS}{cmsy}{m}{n}
\DeclareSymbolFontAlphabet{\mathcal}{usualmathcal}

\fancypagestyle{SPstyle}{
\fancyhf{}
\lhead{\colorbox{scipostblue}{\bf \color{white} ~SciPost Physics }}
\rhead{{\bf \color{scipostdeepblue} ~Submission }}

\fancyfoot[C]{\textbf{\thepage}}
}

\usepackage{tabularray}
\usepackage{bm}
\usepackage{slashed}
\usepackage{enumitem}
\usepackage{physics}
\usepackage{subcaption}
\usepackage{booktabs}

\newcommand{\mbf}[1]{\mathbf{#1}}

\definecolor{colone}{HTML}{fec4de}
\definecolor{coltwo}{HTML}{cbf0f6}
\definecolor{colthree}{HTML}{c7ece0}

\newcommand{\SU}{\mathrm{SU}}

\newcommand{\eqnref}[1]{Eq.~\eqref{#1}}
\newcommand{\figref}[1]{Fig.~\ref{#1}}

\newcommand{\secref}[1]{Sec.~\ref{#1}}
\newcommand{\appref}[1]{Appendix~\ref{#1}}

\newcommand{\aop}{a^{\phantom{\dagger}}}
\newcommand{\adagger}{a^\dagger}
\newcommand{\bop}{b^{\phantom{\dagger}}}
\newcommand{\bdagger}{b^\dagger}
\newcommand{\nop}{n^{\phantom{\dagger}}}
\newcommand{\sigmaop}{\sigma^{\phantom{\dagger}}}
\newcommand{\sigmadagger}{\sigma^{\dagger}}

\newcommand{\sweight}{\mathcal{W}}

\begin{document}

\pagestyle{SPstyle}

\begin{center}{\Large \textbf{\color{scipostdeepblue}{
Momentum-Space Entanglement Signatures and Spinon Breakdown in the $\mbf{J_1}$--$\mbf{J_2}$ Zig-Zag Heisenberg Chain 
}}}\end{center}

\begin{center}\textbf{
Tom~Oeffner\,\orcidlink{0009-0008-9121-9417}\textsuperscript{1,$\ast$},
Ludwig~Bordfeldt\,\orcidlink{0009-0009-4759-6335}\textsuperscript{1,$\ast$},
Andreas~Feuerpfeil\,\orcidlink{0009-0001-0436-0332}\textsuperscript{1,2,$\ast$,$\dagger$},
Lukas~Elter\,\orcidlink{0009-0004-3740-2983}\textsuperscript{1},
Tobias~Helbig\,\orcidlink{0000-0003-1894-0183}\textsuperscript{1,3},
Tobias~Hofmann\,\orcidlink{0000-0002-1888-9464}\textsuperscript{1},
Martin~Greiter\,\orcidlink{0000-0003-2008-4013}\textsuperscript{1},
Ronny~Thomale\,\orcidlink{0000-0002-3979-8836}\textsuperscript{1}
}\end{center}

\begin{center}
\textsuperscript{\bf 1}Institut für Theoretische Physik und Astrophysik and W\"urzburg-Dresden Cluster of Excellence ctd.qmat, Universit\"at W\"urzburg, 97074 W\"urzburg, Germany
\\
\textsuperscript{\bf 2}Center for Computational Quantum Physics, Flatiron Institute, 162 5th Avenue, New York, NY 10010, USA
\\
\textsuperscript{\bf 3}Leinweber Institute for Theoretical Physics, Stanford University, Stanford, CA 94305, USA

{\small $\ast$ These authors contributed equally}\\%
$\dagger$ \href{mailto:andreas.feuerpfeil@uni-wuerzburg.de}{\small andreas.feuerpfeil@uni-wuerzburg.de}
\end{center}

\section*{\color{scipostdeepblue}{Abstract}}
\textbf{\boldmath{%
We investigate the resilience of spinon quasiparticles in the  $J_1$--$J_2$ zig-zag spin chain ($J_2>0$) from the viewpoint of momentum-space entanglement. For small $J_2$, we show that deconfined spinons survive well past the liquid-dimer transition before eventually collapsing towards the Majumdar-Ghosh point. In the highly frustrated zig-zag regime ($J_2 \gg |J_1|$), we model the system as two coupled Heisenberg chains and by Fourier transforming each subchain individually, a framework we dub the double-spinon description. While continuum field theories predict that this decoupled phase is strictly unstable to any finite inter-chain coupling, our analysis reveals that the double-spinon description remains robust over an extensive parameter regime. Notably, we find a stark asymmetry in spinon stability reflecting the underlying renormalization group flow: ferromagnetic coupling ($J_{1} < 0$) is marginally irrelevant and sustains fractionalization deep into the spiral phase, whereas antiferromagnetic coupling ($J_{1} > 0$) is marginally relevant and drives confinement much earlier. The ultimate breakdown of this fractionalized description is driven by a continuum of inter-chain excitations which manifests itself as a sharp ground-state momentum shift distinct from macroscopic thermodynamic phase boundaries. Our results establish momentum cut entanglement analysis as a tool to trace the quasiparticle resilience of spinons, as we show that treating the zig-zag Heisenberg chain as two coupled SU(2)$_1$ Wess-Zumino-Witten models provides a theoretical framework for strongly frustrated quantum magnets applicable beyond the decoupled limit.
}}

\vspace{\baselineskip}

\noindent\textcolor{white!90!black}{%
\fbox{\parbox{0.975\linewidth}{%
\textcolor{white!40!black}{\begin{tabular}{lr}%
  \begin{minipage}{0.6\textwidth}%
    {\small Copyright attribution to authors. \newline
    This work is a submission to SciPost Physics. \newline
    License information to appear upon publication. \newline
    Publication information to appear upon publication.}
  \end{minipage} & \begin{minipage}{0.4\textwidth}
    {\small Received Date \newline Accepted Date \newline Published Date}%
  \end{minipage}
\end{tabular}}
}}
}


\vspace{10pt}
\noindent\rule{\textwidth}{1pt}
\tableofcontents
\noindent\rule{\textwidth}{1pt}
\vspace{10pt}

\noindent
\section{Introduction}
\label{sec:introduction}

\begin{figure}[t]
    \centering
    \includegraphics[width=0.8\linewidth]{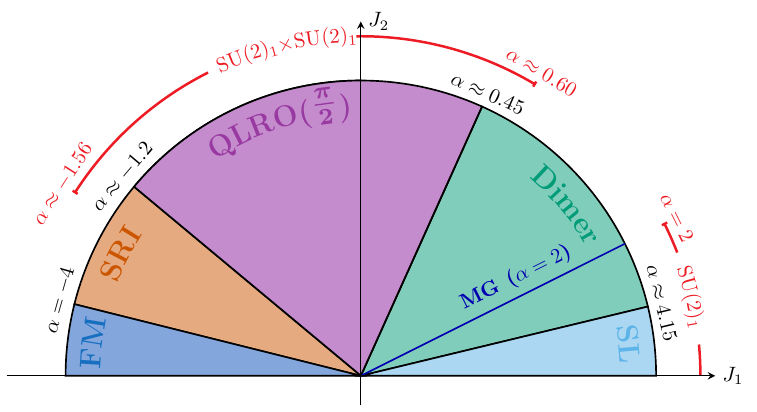}
    \caption{Quantum phase diagram of the $J_1$--$J_2$ spin-$1/2$ chain parameterized by the inverse frustration ratio $\alpha = J_1/J_2$. The model hosts a variety of ground states: a gapless Heisenberg spin liquid (SL), a gapped dimer phase breaking translation symmetry~\cite{okamoto_1992_fluiddimer}, a quasi-long-range order phase [QLRO($\pi/2$)]~\cite{kumar_2015_level,soos_2016_numerical}, a short-range incommensurate (SRI) spiral phase~\cite{kumar_2015_level,soos_2016_numerical}, and a ferromagnetic (FM) phase~\cite{White_1996,sandvik_2010_computational}. In this work, we argue that fractionalized spinon descriptions (indicated by the outer red arcs) extend significantly beyond their non-interacting limits. The single-chain $\SU(2)_1$ description survives past the SL--dimer transition ($\alpha \approx 4.15$) and persists up to the Majumdar-Ghosh (MG) point at $\alpha = 2$. Furthermore, the double-spinon description of the spin chain as two weakly interacting chains [$\SU(2)_1 \times \SU(2)_1$], natural for large $J_2$, remains robust well beyond the field theory prediction, that QLRO($\pi/2$) is constraint to $\alpha = 0$~\cite{White_1996,allen_1997_nonabelian,dmitriev_2008_weakly,sudan_2009_emergent,sirker_2011_j1j2,furukawa_2012_groundstate}, reaching into the SRI phase down to $\alpha \approx -1.56$ and into the dimer phase up to $\alpha \approx 0.60$. The breakdown of the deconfined spinon description is marked by an abrupt shift in the ground-state momentum character. The critical values of $\alpha$ are obtained by a linear extrapolation to the thermodynamic~limit.}
    \label{fig:j1_j2_phasediagram}
\end{figure}

The concept of fractionalization, wherein local collective excitations carry a fraction of the quantum numbers associated with the underlying particles, constitutes a cornerstone of modern quantum condensed matter physics. In one-dimensional quantum magnets, the canonical manifestation of this phenome\-non is the spinon---a spin-$1/2$ excitation emerging from the spin-$1$ magnons of a classically ordered state. The $J_1$--$J_2$ spin-$1/2$ chain has long served as a fundamental paradigm for exploring the interplay between frustration, quantum fluctuations, and these fractionalized excitations. Driven by the competition between nearest-neighbor (NN) and next-nearest-neighbor (NNN) interactions $J_1$ and $J_2$, the system is described by the Hamiltonian
\begin{equation}\label{main_eq: real-space Hamiltonian}
    H = J_1 \sum_{i} \mathbf{S}_{i} \cdot \mathbf{S}_{i+1} + J_2 \sum_{i} \mathbf{S}_i \cdot \mathbf{S}_{i+2}\,.
\end{equation}
While recent works~\cite{zirnbauer_2024_infrared,mcroberts_2025_transition,azad_2025_generalized} extensively investigated generalized phase boundaries and infrared transitions in models featuring ferromagnetic NNN couplings ($J_2 < 0$), we focus our analysis on the antiferromagnetic sector ($J_2 > 0$), where frustation can occur. Depending on the exact ratio of these couplings, the model hosts a rich phase diagram encompassing gapless Luttinger liquid, gapped dimerized, and incommensurate spiral phases (see \figref{fig:j1_j2_phasediagram})~\cite{White_1996,sandvik_2010_computational,kumar_2015_level,soos_2016_numerical}.

While the exact nature of these fractionalized excitations is strictly known at integrable points, their robustness away from these ideal limits remains an open question. To establish a baseline, we first consider the single-chain limit ($J_2 = 0$), which is in the same universality class as the Haldane--Shastry spin chain~\cite{haldane_1988_exact,shastry_1988_exact}, the exact lattice realization of the $\SU(2)_1$ Wess-Zumino-Witten (WZW) model~\cite{wess_1971_consequences,novikov_1982_hamiltonian,witten_1984_nonabelian,knizhnik_1984_current}. In the Haldane--Shastry model, spinons only interact through their exchange statistics and are protected by an infinite entanglement gap (EG) in a momentum entanglement cut~\cite{Thomale2_2010,Thomale_2010,Lundgren_2012,Lundgren_2014,Lundgren_2016,Dora_2016,ibanez-berganza_2016_fourierspace,Bayat2022EntanglementSpinChains}. In this work, we provide evidence that the spinon description in the $J_1 > J_2$ regime survives well past the liquid-dimer transition at $J_2 \approx 0.241J_1$. Within this $J_1$--$J_2$ spin fluid regime where we find spinons to be the adequate quasiparticle description, the point exhibiting the highest spinon weight is in close proximity to the maximum momentum cut EG. It is also the point where the ground state most strongly resembles that of the Haldane--Shastry model. Upon further increasing $J_1-J_2$ the fractionalized spinon description eventually collapses towards the Majumdar-Ghosh point~\cite{majumdar_1969_nextnearestneighbora,majumdar_1969_nextnearestneighbor}.

Building upon this foundation deriving from the single-Heisenberg chain regime, attention has recently shifted to the $J_1$--$J_2$ model in the zig-zag limit ($J_2 \gg |J_1|$), where the ferromagnetic regime ($J_1 < 0$) has garnered particular interest due to its relevance to Cu(II) chains in cupric oxides~\cite{hase_2004_magnetic,enderle_2005_quantum,lu_2006_zigzag,drechsler_2007_frustrated,banks_2007_high,dutton_2012_quantum,soos_2016_numerical}. Here, the system is naturally conceptualized as two interacting copies of a Heisenberg chain spin fluid regime. In the decoupled limit ($J_1 = 0$), the ground state is a direct product of two such states, inherently described by deconfined spinon degrees of freedom. A fundamental claim of most field theories studied to date~\cite{White_1996,allen_1997_nonabelian,dmitriev_2008_weakly,sudan_2009_emergent,sirker_2011_j1j2,furukawa_2012_groundstate} is that the quasi-long-range order phase at wavevector $\pi/2$ [QLRO($\pi/2$)] were strictly limited to this $J_1=0$ point, and hence extremely sensitive even to perturbative deviations. As discussed in previous analytical studies~\cite{White_1996,allen_1997_nonabelian}, the system is expected to be gapless for $J_1<0$ (even though the expectation of a strongly reduced gap has also been reported~\cite{Itoi_2001}) and to possess an exponential gap for $J_1>0$, where the structure of the excitations was generally deemed inconsistent with the decoupled limit. 

Nevertheless, numerical evidence suggests that the QLRO($\pi/2$) phase remains stable, persisting over a finite range of the exchange interaction $J_1$~\cite{soos_2016_numerical}. To contribute to resolving the tentative inconsistency between numerics and quantum field theory, we model the zig-zag spin chain as two interacting Heisenberg spin liquids and Fourier transform each of the two subchains into an effective spinon basis~\cite{Thomale2_2010}, which we dub the double-spinon description. By considering the momentum-space entanglement spectrum (ES) and the associated EG, we present convincing evidence for the persistence of spinons well beyond the decoupled limit. We show that this fractionalized description remains valid up to values of $J_1/J_2$, ranging from roughly $-1.56$ to $0.60$ subject to systematic errors from finite size extrapolation. Because established numerical studies~\cite{kumar_2015_level,soos_2016_numerical} place the macroscopic boundaries of the QLRO($\pi/2$) phase at $J_1/J_2 \approx -1.2$ and $0.45$, which is fully contained within our extracted limits of the double-spinon regime. 

Crucially, our momentum entanglement cuts reveal that the double-spinon description does not succumb immediately to the onset of inter-chain coupling, but survives into the nonzero $J_1$ regime. We demonstrate that the eventual closure of the EG is not driven by the renormalization of intra-chain entanglement levels but is rather dictated by a descending continuum of inter-chain spinon excitations. For both signs of $J_1$, this breakdown is associated with a sharp change in the momentum character of the ground state, characterized by a strong admixture of different momentum sectors between the two chains. While our numerical data are restricted to system sizes that are multiples of four to maintain a spin-singlet ground state on each sub-chain, the smooth convergence of our results suggests that this association remains a robust feature in the thermodynamic limit. The frustration ratios $J_1/J_2$ characterizing this momentum transition are extrapolated to the thermodynamic limit via a real-space scaling analysis. This represents a distinct boundary where the non-local properties captured within the ES are abruptly scrambled. The observation that such a momentum entanglement transition occurs shifted from the thermodynamic phase boundaries is consistent with previous studies~\cite{Thomale2_2010,Lundgren_2014}.

Despite sharing this momentum-shift mechanism, we observe a stark asymmetry in the stability of the spinon description depending on the sign of the coupling. For ferromagnetic inter-chain coupling ($J_1 < 0$), the double-spinon description remains robust, surviving well past the thermodynamic phase boundary---located at $J_1/J_2 \approx -1.2$ according to density matrix renormalization group (DMRG) calculations~\cite{soos_2016_numerical}---and into the spiral phase to $J_1\approx -1.56J_2$. Conversely, for antiferromagnetic coupling ($J_1 > 0$), our momentum EG analysis reveals that the spinon confinement transition occurs inside the dimer phase around $J_1\approx 0.60J_2$, distinctly separated from the DMRG phase boundary at $J_1/J_2 \approx 0.45$~\cite{soos_2016_numerical}. We directly connect this breakdown to the shift in the ground-state momentum, which we extract via our real-space scaling analysis.

This observed asymmetry can be elegantly reconciled through the lens of conformal field theory (CFT). Employing methods pioneered by Ian Affleck~\cite{Affleck_1985,Affleck_1986,Affleck_1987,affleck_1987_critical,affleck_1989_quantum}, we shortly revisit the field-theoretic analysis of the zig-zag chain by treating the system as two coupled $\mathrm{SU}(2)_1$ WZW models. Continuum field theory predicts that this description is destroyed by any finite inter-chain coupling---implying that fractionalized spinons should not survive beyond $J_1=0$~\cite{White_1996,allen_1997_nonabelian,dmitriev_2008_weakly,sudan_2009_emergent,sirker_2011_j1j2,furukawa_2012_groundstate}. As we demonstrate through entanglement spectra~\cite{Bayat2022EntanglementSpinChains}, our results challenge this strict confinement boundary, showing that the double-spinon description persists well into the finite $J_1$ regime. However, the field theory captures the underlying dynamics of this breakdown through its renormalization group (RG) flow. For antiferromagnetic inter-chain coupling ($J_1 > 0$), the inter-chain interaction acts as a marginally relevant operator that generates an exponentially small gap, actively driving the instability~\cite{White_1996,allen_1997_nonabelian}. In contrast, ferromagnetic coupling ($J_1 < 0$) is marginally irrelevant, meaning this specific perturbative instability is absent. This RG prediction mirrors the stark asymmetry we observe: a rapid breakdown inside the dimer phase for $J_1 > 0$, contrasted with an extended stability of the double-spinon description deep into the spiral phase for $J_1 < 0$.

Ultimately, our study intends to establish that framing the zig-zag chain as two interacting copies of the Heisenberg spin fluid phase provides an accurate theoretical framework over an extensive parameter regime. By demonstrating the robustness of this double-spinon picture, our results open a potential avenue for deriving rigorous field-theoretic descriptions---perspectively also for the hiterto largely elusive spiral phase---starting directly from these low-energy, fractionalized degrees of freedom.

\section{Spinon description of the \texorpdfstring{$\mbf{J_1}$--$\mbf{J_2}$}{J1--J2} Heisenberg spin chain}

\subsection{Model and details of a momentum-space partition}\label{sec: model and details of a momentum-space partition}
We consider the one-dimensional $J_1$--$J_2$ Heisenberg spin-$1/2$ model, governed by \eqnref{main_eq: real-space Hamiltonian}, on an even-membered chain of $L$ spins with periodic boundary conditions. 
To isolate the fractionalized excitations and probe their entanglement structure, we construct a momentum-space representation of the Hilbert space (see \appref{appendix: transformed hamiltonian in the single-spinon formalism} for a detailed derivation). In particular, we map the spin operators onto hard-core bosons via $S_i^+ = b_i^\dagger$ and $S_i^z = \bdagger_i \bop_i - 1/2$~\cite{matsubara_1956_lattice}. 
The additional hard-core constraint $ V(\bdagger_i \bop_i- 1/2)^2$ with~$V \to \infty$, which prevents double occupancy of the bosonic modes, acts as an infinite  on-site repulsion in real space. Crucially, this local constraint scatters particles across all momenta, rendering it a primary source of entanglement when the system is partitioned in momentum space. In our numerical simulations, we employ $V=10^3$. By transforming to momentum space via
\begin{equation}\label{main_eq: fourier transform of bosonic operators}
    b_j = \frac{1}{\sqrt{L}}\sum_k e^{i k j} b_k\,,
\end{equation}
where the crystal momentum is $k = \frac{2\pi}{L}n$ with $n \in \{-L/2+1, \dots, L/2\}$, the basis states can be expressed in terms of bosonic occupation numbers $n_k$ for each mode $k$. The system possesses two exact quantum numbers: the total particle number $N = \sum_k n_k$ (corresponding to the conserved total magnetization $S^z$) and the total crystal momentum $K_c = K \pmod{2\pi}$, where $K = \sum_k n_k k$ is the unfolded total momentum. 

Since $K_c$ is strictly conserved, the ground-state wave function generally exhibits finite support across multiple total momentum sectors $K=K_c\pm  2\pi j$ with $j\in\mathbb{N}_0$. For states with compact support, however, the spectral weight is overwhelmingly concentrated within a single momentum sector $K$, rendering it an approximate quantum number. In this context, we formally define the (highest) spinon weight as the squared overlap of the ground state $\ket{\Psi_0}$ with that sector, i.e.,
\begin{align} \label{main_eq: spinon weight}
    \sweight \equiv \max_{K} \sweight(K)\, \qquad \mathrm{with} \qquad \sweight(K) = \bra{\Psi_0} \mathcal{P}_K \ket{\Psi_0}\,,
\end{align}
where $\mathcal{P}_K$ projects onto the subspace of total momentum $K$. We track the spinon weight as a measure of spinon quality to characterize the confinement, enabling a direct comparison between the ground-state wave functions and the exact momentum sectors of the Haldane--Shastry model.

In the following, we use exact diagonalization (ED) to analyze the momentum-space ES of the ground state, which is presumed to be a total spin singlet with $N=L/2$ and $K_c=0$ or $\pi$. To this end, we introduce a bipartition of the momentum orbitals into two regions, $A=\{k|k>0\}$ and $B=\{k|k\le 0\}$, constraining both the number of particles $N_A+N_B=N$ and the total crystal momenta $(K_A + K_B)\pmod{2\pi}=K_c$. In the case where the total momentum $K$ is approximately conserved, we partition the system with respect to $K_A+K_B=K$. Because we choose a cut that respects the translational symmetry of the system, the reduced density matrix $\rho_A$ inherits this symmetry. In particular, $\rho_A$ is block-diagonal in the total crystal momentum $M_{A,c} = M_A \pmod{2\pi}$ allowing us to plot the entanglement levels $\xi$ (which are obtained through the eigenvalues of $\rho_A$) with respect to the quantum numbers $K_{A,c}$ or $K_A$.

If the system admits a CFT description, the ES is expected to mimic the characteristic state counting of the CFT within a set of low-lying universal entanglement levels. These levels are separated from the generic entanglement weights by an EG~\cite{Li_2008,Thomale2_2010}, which we define as
\begin{align}\label{main_eq: direct EG}
    \Delta \equiv \min_{K_{A,c}} \{  \min_\xi \xi^\mathrm{generic}_{K_{A,c}} - \max_\xi \xi^\mathrm{universal}_{K_{A,c}} \}\,,
\end{align}
i.e., the minimal direct gap over all sectors $K_{A,c}$. While this definition is primarily used throughout this work (see \figref{fig_ssES_a}), one may alternatively introduce an indirect gap globally across all momentum sectors,
\begin{align}\label{main_eq: indirect EG}
    \bar{\Delta} \equiv \min_\xi \xi^\mathrm{generic} - \max_\xi \ \xi^\mathrm{universal}\,.
\end{align}
In our analysis, however, the spectral flow of entanglement levels extends coherently across all sectors $K_{A,c}$, which is why both definitions yield nearly identical gap values (see \appref{appendix: measures of spinon confinement}). Due to the computational complexity of ED, we are restricted to system sizes up to $L=22$, which corresponds to matrices containing up to several billions of scattering elements.

\subsection{Conservation of spinon momentum and entanglement gap}

Previous work by one of us~\cite{Thomale2_2010} demonstrated that the momentum-space ES at the liquid-dimer transition in the antiferromagnetic regime $(J_1, J_2 >0)$ detects the underlying topological and conformal order of the Heisenberg chain---that is, the critical spin fluid phase realizes the SU$(2)_1$ WZW model with fractionalized spinons as elementary excitations~\cite{Abrikosov-1965,Wen-1991}. A central conjecture emerging from\cite{Thomale2_2010} is that an EG separates a non-universal, model-dependent part of the ES from a universal low-lying part, which encodes a fingerprint of the associated CFT~\cite{Li_2008}. Notably, this fundamental structure typically remains hidden in more conventional real-space partitions and is only revealed through a local cut in momentum space~\cite{Thomale2_2010,PhysRevLett.110.046806}.

\begin{figure}[t]
    \centering
    \includegraphics[width=1.0\linewidth]{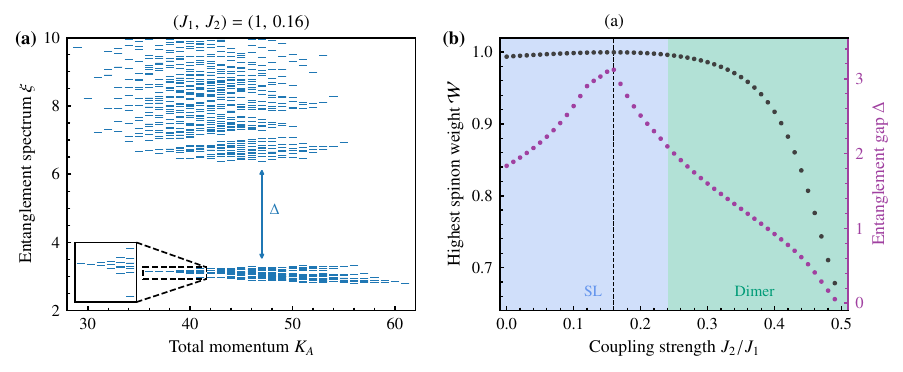}
    \caption{(a) Momentum entanglement spectrum for $L=22$ sites, with a cut region containing $N_A=6$ particles. The SU$(2)_1$ low-lying levels are well separated from a non-universal part by an entanglement gap $\Delta$ which persists in the thermodynamic limit~\cite{Thomale2_2010}. (b)~Maximum spinon weight $\sweight$ in the $K=\pi L/2$ sector and gap $\Delta$ versus frustration coupling $J_2/J_1 >0$. The spinons proliferate deep inside the dimer phase until the gap $\Delta$ finally collapses at the Majumdar-Ghosh point. Both $\sweight$ and $\Delta$ are highest for finite $J_2$ (dashed line), marking the parameter region of the system being closest to the Haldane--Shastry model, as indicated by their ground-state overlap in \appref{appendix: measures of spinon confinement}. }

    \label{fig_ssES}
    \phantomsubcaption\label{fig_ssES_a}
    \phantomsubcaption\label{fig_ssES_b}
\end{figure}

A representative momentum-space ES within the spin-liquid phase is presented in \figref{fig_ssES_a} for a total momentum cut with $L=22$ and $N_A=5$. We observe a clear separation between low-lying entanglement levels and higher levels, separated by an EG that remains finite in the thermodynamic limit when assuming a linear scaling with $1/L$~\cite{Thomale2_2010}. The low-lying CFT levels follow the state counting $1,1,2,3,5,7\dots$, which is equivalent to that of a free U(1) boson. This is no coincidence, as the Heisenberg chain lies in the same universality class as the Haldane--Shastry chain~\cite{haldane_1988_exact,shastry_1988_exact}, whose gapless spinon excitations host the same counting as the edge excitations of the fractional quantum Hall effect. This can be explicitly seen in terms of the Fourier-transformed Haldane--Shastry wave function, which yields the same weights of monomials as the Laughlin state~\cite{Thomale2_2010}.
Its edge theory is a chiral U(1) boson CFT whose level counting corresponds to the number of partitions of its momentum above the ground state. Because this field theory is equivalent to the SU$(2)_1$ WZW model~\cite{Affleck_1986, Sonnenschein_1988}, there is a one-to-one correspondence between the low-energy excitations of the bulk and the momentum-space ES. 

Although the state counting of the low-lying levels in the ES is identical for the ground states of the Haldane--Shastry and Heisenberg models and one can interpolate between them without the EG closing~\cite{Thomale_2010, Thomale2_2010}, there is an important difference: toward the Heisenberg point, the spinon excitations interact beyond their fractional statistics, dressing the state in two fundamental ways. First, while the ES for the Haldane--Shastry wave function consists purely of the SU$(2)_1$ low-lying levels with an infinite EG, the onset of interactions between spinons introduces logarithmic CFT corrections. These populate high-entanglement levels that render the EG finite, but leave the low-energy excitations of the bulk invariant. Second, unlike the Haldane--Shastry model where the total spinon momentum $K=\pi N$ is an exact quantum number, the spin-liquid wave function generates a Gaussian admixture of components with $K=\pi N\pm 2\pi j$ for $j\in \mathbb{N}_0$. 

However, because a large percentage of the spectral weight of the Heisenberg antiferromagnet is still located within the highest weight sector $K=\pi N$, the unfolded total momentum serves as a highly useful approximate quantum number. In this context, we suggest the spinon weight as a diagnostic criterion for the validity of a deconfined spinon description in quantum spin chains. As presented in \figref{fig_ssES_b}, the spinon weight remains remarkably high $(>0.993)$ throughout the spin-liquid phase, attaining a maximum at finite $J_2$. Importantly, this point marks the regime in which the ground state most closely resembles that of the Haldane--Shastry chain (see \appref{appendix: measures of spinon confinement}), and, within the limits of our numerical resolution, coincides with the largest direct EG. Indeed, consistent with the Sine-Gordon description of the frustrated Heisenberg spin chain~\cite{Haldane_1982}, the leading marginal operator is conjectured to vanish at this point~\cite{Thomale2_2010}, raising the higher levels of the ES and thereby maximizing the EG. Upon tuning the coupling $J_2$ above this critical value (where the Sine-Gordon cosine term changes sign), both the spinon weight and the EG gradually decrease before ultimately collapsing at the Majumdar-Ghosh point. This demonstrates that the deconfined spinon description remains robust even beyond the dimerization transition at $J_2=0.241J_1$~\cite{Eggert_1996}, protected by a high spinon weight and a finite EG.

Our conclusions depend mildly on whether we define the minimal EG as a direct gap $\Delta$ within a given sector $(K_{A,c}, N_A)$ or as a indirect gap $\bar{\Delta}$ over all $K_{A,c}$. When considering the indirect gap $\bar{\Delta}$, the maximum spinon weight lies slightly closer to the dimerization transition than the largest EG and instead, within our numerical resolution, coincides with the maximum overlap with the Haldane--Shastry ground state (see \appref{appendix: measures of spinon confinement}). While a strict one-to-one correspondence is not expected, the close agreement between these quantities provides a consistent, robust diagnostic of spinon deconfinement and validates the spinon description.

\subsection{Breakdown of the spinon description}
The limit of weak NN coupling, $|J_1| \ll J_2$, as well as the short-range incommensurate (SRI) phase---also called the spiral phase---in the frustrated ferromagnetic-antiferro\-magnetic regime, has not been a primary focus of field theory studies for many years. Despite considerable recent interest, particularly regarding the presence of an exponentially small energy gap~\cite{White_1996,allen_1997_nonabelian,Itoi_2001}, no low-energy CFT describing this phase has been firmly identified. In this context, we position the momentum-space ES as a highly promising tool to probe the low-energy excitations of the bulk that characterize the principal nature of excitations of the system.

\begin{figure}[t]
    \centering
    \includegraphics[width=1.0\linewidth]{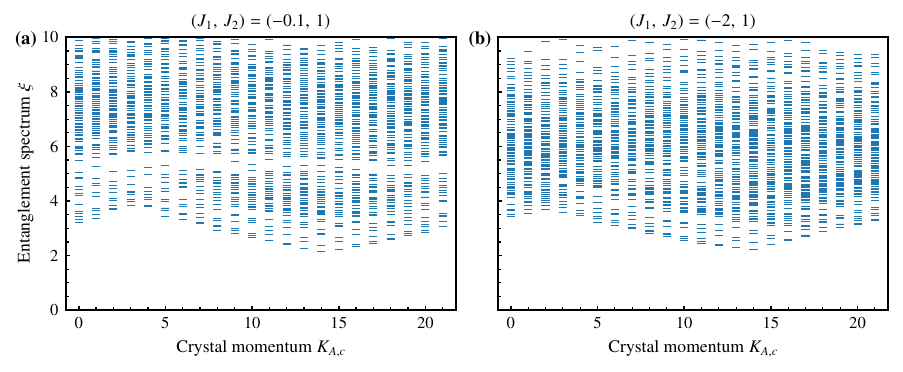}
    \caption{Entanglement spectrum for $L=22$ sites, with a cut region containing $N_A=6$ particles. Inside the QLRO$(\pi/2)$ (a) and the spiral phase (b), the entanglement levels form a continuum of states without any finite entanglement gap, indicating that a description in terms of single-chain spinons is not appropriate. }

    \label{fig_ssSpiral}
    \phantomsubcaption\label{fig_ssSpiral_a}
    \phantomsubcaption\label{fig_ssSpiral_b}
\end{figure}

Because the spiral phase is a macroscopic singlet state and is generally expected to be gapless or to possess a highly suppressed gap, one might intuitively expect that the phase lends itself to a description based on fractionalized spinon degrees of freedom. However, as shown in \figref{fig_ssSpiral} for exemplary data points in the QLRO($\pi/2$) and spiral phase, our analysis of the ES in this extended parameter regime reveals a dense continuum of entanglement levels without any finite EG. Because we do not observe a universal low-lying part of the ES protected by such a gap, we can effectively rule out single-chain spinons as the emergent quasiparticles of the spiral or the QLRO($\pi/2$) phase. This conclusion is further corroborated by the ground-state wave function, which exhibits strongly distributed weight across several momentum sectors, with the maximum spinon weight remaining unremarkably small ($\sweight < 0.3$) throughout the considered parameter range. This qualitative behavior is likewise observed in the dimer phase beyond the Majumdar-Ghosh point.

Consequently, a deconfined single-spinon description fails to capture the non-trivial fundamental structure in these phases. This breakdown naturally prompts a conceptual question: given that the spiral phase extends to considerable values of NNN coupling, is the interpretation of the zig-zag chain as a single, fundamentally one-dimensional chain incorrect? We argue that in this strongly frustrated regime, the single-chain paradigm is indeed inadequate. Instead, the system must be treated as two weakly coupled, interacting copies of a Heisenberg spin fluid state. In the following, we demonstrate that this double-spinon description successfully captures the underlying physics.

\section{Zig-zag limit of two weakly interacting chains (\texorpdfstring{$\mbf{J_1 \to 0}$}{J1 to 0})}
\label{sec:decoupled_limit}

The $J_1=0$ limit of the zig-zag chain corresponds to two completely decoupled chains residing on the sublattices of odd- and even-numbered sites. Governed exclusively by the intra-chain Heisenberg interaction $J_2 > 0$, both sublattices are exactly described as gapless Heisenberg spin liquids. Having established the failure of the single-chain description in the strongly frustrated regime, we now investigate how these decoupled spin liquids break down upon the introduction of the inter-chain perturbation $J_1$, utilizing the same entanglement measures.

\subsection{Bosonization and Abrikosov-fermion description}

To establish a theoretical baseline for the zig-zag chain in the weakly coupled regime $,|J_1| \ll J_2$, we turn to the non-Abelian bosonization framework. Following the pioneering methodology developed by Affleck~\cite{Affleck_1985,Affleck_1986,Affleck_1987,affleck_1987_critical,affleck_1989_quantum}, the decoupled limit is described by two independent $\mathrm{SU}(2)_1$ WZW models. 

Upon expressing the spin operators in terms of Abrikosov fermions and switching to a current algebra representation, the inter-chain perturbation $J_1$ introduces interactions between the left- and right-moving currents of the two subchains. As established by White and Affleck~\cite{White_1996} and Allen \textit{et al.}~\cite{allen_1997_nonabelian}, the zig-zag geometry results in a profound departure from the standard two-leg spin ladder. In a spin ladder, the inter-chain ``rung'' coupling acts as a strongly relevant perturbation. It generates a spin gap that scales linearly with the coupling strength---up to logarithmic corrections---regardless of the coupling's sign.

In stark contrast, the staggered geometry of the zig-zag chain causes the leading relevant inter-chain operators to cancel out dynamically. The one-loop RG analysis of the remaining marginal current-current interactions~\cite{ludwig_2003_4loop} reveals a striking asymmetry: the effective inter-chain coupling is marginally relevant for $J_1 > 0$ and marginally irrelevant for $J_1 < 0$. Consequently, for antiferromagnetic coupling, the system dynamically generates a gap that scales exponentially with the inverse perturbation, $\Delta \propto \exp(-\text{const} \cdot J_2/J_1)$. For ferromagnetic coupling, the marginally irrelevant nature of the interaction prevents this gap generation, leaving the system gapless (or with an infinitesimally small gap as proposed in~\cite{Itoi_2001}) and stabilizing the extended QLRO($\pi/2$) phase. This field-theoretic asymmetry anticipates the delayed confinement and extended spinon stability we observe numerically for $J_1 < 0$ in the next section.

\subsection{Entanglement spectroscopy of the double-spinon description}

\begin{figure}
    \centering
    \includegraphics[width=1.0\linewidth]{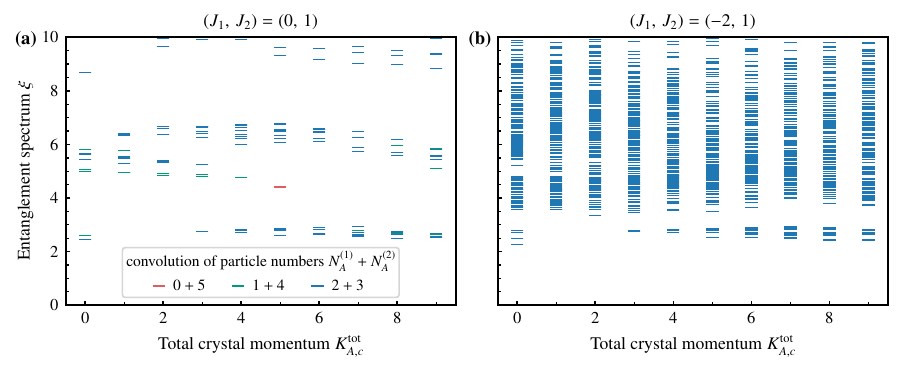}
    \caption{(a) Entanglement spectrum in the decoupled limit for $L_c=10$ unit cells, with a cut containing $N_A=5$ particles. The spectrum is given by the convolution of two independent $L=10$ Heisenberg spin-chains with nearest-neighbor coupling only. The low-lying entanglement levels are described by two $\SU(2)_1$ WZW models. (b) Entanglement spectrum within the spiral phase. Both the CFT levels as well as the intra-chain Heisenberg levels do not renormalize significantly, whilst a continuum of inter-chain excitations moves down without closing the entanglement gap.}
    \label{fig_dsES}
    \phantomsubcaption\label{fig_dsES_a}
    \phantomsubcaption\label{fig_dsES_b}
\end{figure}

In the limit of two decoupled chains, labeled by $1$ (odd sites) and $2$ (even sites), we enlarge the unit cell of the system to contain two spins. Accordingly, the real-space Hamiltonian defined in \eqnref{main_eq: real-space Hamiltonian} for a periodic chain of $L_c=L/2$ unit cells can be reformulated as
\begin{equation}
    H = J_1\sum_{i=1}^{L_c}\left( \mathbf{S}_i^{(1)} \cdot \mathbf{S}_i^{(2)} + \mathbf{S}_i^{(1)}\cdot \mathbf{S}_{i-1}^{(2)} \right) + J_2 \sum_{i=1}^{L_c} \left( \mathbf{S}_i^{(1)} \cdot \mathbf{S}_{i+1}^{(1)} + \mathbf{S}_i^{(2)} \cdot \mathbf{S}_{i+1}^{(2)}\right)\,.
\end{equation}
To construct a momentum-space representation, we apply the single-chain formalism outlined in \secref{sec: model and details of a momentum-space partition} to each subchain, assigning distinct flavors of hardcore bosons to the odd and even sites (a detailed derivation of this formalism is provided in \appref{appendix: transformed hamiltonian in the double-spinon formalism}). In this language, the two sublattice crystal momenta $K_c^{(i)}=\sum_k n_k^{(i)}k^{(i)}_{\phantom{k}} \pmod{2 \pi}$ with $i = 1,2$ are conserved independently for $J_1=0$, whereas only the total crystal momentum $K_c^\mathrm{tot} = (K_c^{(1)} + K_c^{(2)}) \pmod{2\pi}$ remains an exact quantum number for finite inter-chain coupling. The unfolded total momenta $K^{(i)}$ allow us to define the (highest) spinon weights on each subchain analogous to \eqnref{main_eq: spinon weight}, which, for an even number of unit cells $L_c$, are identical to each other, $\sweight^{(1)}=\sweight^{(2)}\equiv\sweight$.

To probe the non-trivial fundamental structure of the quantum phases emerging from the decoupled limit, we analyze not only the ground state but also the low-energy excitation spectrum. We restrict the wave functions to be total spin singlets with $N=L_c$ and $K_c^\mathrm{tot}=0$, containing both the $K_c=0$ and $\pi$ sectors due to the enlarged unit cell. To apply a joint middle cut in momentum space, the system is partitioned into $A=\{(k^{(1)},k^{(2)}) \mid k^{(1)},k^{(2)}>0\}$ and $B=\{(k^{(1)}, k^{(2)}) \mid k^{(1)}, k^{(2)}\le0\}$, constraining the number of particles and the total crystal momentum of $A$ and $B$, i.e., $N_A + N_B=N$ and $(K_A^\mathrm{tot} + K_B^\mathrm{tot})\pmod{2 \pi} = K_c^\mathrm{tot}$. 

To ensure that each subchain can host a valid spin-singlet ground state, the total number of unit cells $L_c$ must be a multiple of two. This stringent symmetry requirement, combined with the doubling of the momentum-space basis and the emergence of complex-valued matrices in the double-spinon formalism (reflecting the inter-chain coupling $J_1$), heavily restricts our numerical analysis to system sizes up to $L_c=10$, making finite-size scaling highly challenging.

\subsubsection{From the decoupled limit to a finite inter-chain perturbation}

We first consider the ground state of the zig-zag chain in the decoupled limit $J_1=0$. Here, the ground-state wave function factorizes into the tensor product of two Heisenberg spin-liquid states. Because this structure is inherited by the reduced density matrix, the ES is given by a direct convolution of the spectra of the individual subchains (see \figref{fig_dsES_a}). Consequently, the momentum-space ES at $J_1=0$ is characterized by a finite EG and a high spinon weight within the sublattice total momentum sector $K^{(i)} = \pi N/2$. The lowest levels accurately follow the $\mathrm{SU}(2)_1 \times \mathrm{SU}(2)_1$ WZW model, while higher levels correspond to the non-universal intra-chain levels of the Heisenberg spin chain (which is not at the conformal fixed point of the Haldane--Shastry model, where the EG is infinite).

In \figref{fig_dsES_a}, this structure manifests in the ES for a middle cut with $L_c=10$ and $N_A=5$. Because the convolution involves all partitions of $N_A = N_A^{(1)} + N_A^{(2)}$ that are invariant under the exchange of the subchains, the resulting entanglement levels are two-fold degenerate. Upon introducing a finite perturbation $J_1$, this degeneracy is lifted (see \figref{fig_dsES_b}), reflecting the failure of the exact tensor product structure. Physically, this arises from inter-flavor interactions between spinons on the two subchains, which dress the product-like wave function. However, as seen in \figref{fig_dsES_b} for a data point deep within the spiral phase, the ES cleanly rearranges while remaining protected by a finite EG and a high spinon weight. This indicates that the low-energy structure of the perturbed system can still be well-described by two coupled $\mathrm{SU}(2)_1$ WZW models. In the following, we demonstrate that this framework remains robust over a broad parameter regime for both signs of $J_1$ and breaks down only when the EG collapses due to a continuum of inter-chain excitations.

\subsubsection{Breakdown of the double-spinon description}
\begin{figure}
    \centering
    \includegraphics[width=1.0\linewidth]{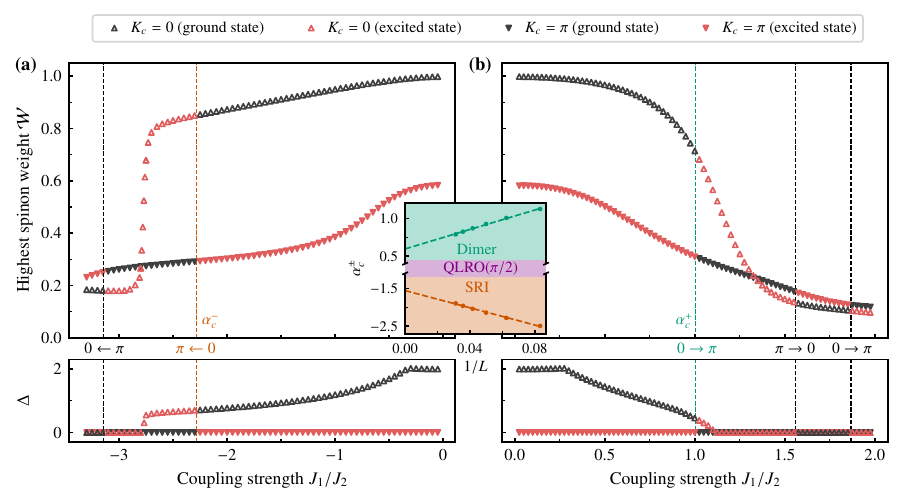}
    \caption{Highest spinon weight $\mathcal{W}$ and entanglement gap $\Delta$ as a function of $J_1/J_2$ for the low-energy $K_c=0$ and $\pi$ state of a system with $L_c=8$ unit cells and $N_A=4$ particles. The ground state is marked in black. We observe that $\mathcal{W}$ and $\Delta$ change continuously and break down only deep inside the frustrated regime $J_1<0$ (a) and $J_1>0$ (b). The points $\alpha_c^{\pm}$ of spinon confinement within the ground-state, which we identify as the gap closing for both signs of $J_1$, is given by the change of momentum character from $0$ to $\pi$. This entanglement transition does not coincide with the phase transition boundaries of the spiral (SRI) and dimer phase, as shown by the linear extrapolation of this level crossing to the thermodynamic limit (see inset). Assuming a~$1/L$ scaling, we obtain the values $\alpha_c^-=-(1.556\pm0.018)$ and $\alpha_c^+=0.596\pm0.011$.}
    \label{fig_dsEG}
\end{figure}

To probe the stability of the double-spinon framework away from the decoupled limit, we track the evolution of the EG and the momentum-resolved spinon weight of the low-energy states at $K_c=0$ and $\pi$ as a function of the inverse frustration ratio $\alpha = J_1/J_2$, as presented in \figref{fig_dsEG}. Note that due to the convergence limits of the Lanczos algorithm for excited states, we use $L_c=8$ unit cells here instead of $L_c=10$ as in \figref{fig_dsES}. For completeness, the spectral flow of entanglement levels from the decoupled limit towards the dimer and spiral phase is shown in \appref{appendix: evolution of entanglement spectra}.

Crucially, we observe that spinon confinement sets in only at considerably large values of $|J_1|$. This breakdown is marked by a sharp closure of the EG and a discontinuous collapse of the maximum spinon weight due to inter-chain levels, rather than a smooth renormalization of intra-chain excitations. Though evolving continuously within each momentum sector $K_c$, these non-analytic features in the ES are driven by an abrupt shift in the ground-state momentum character between $0$ and $\pi$. By tracking these finite-size crossings $\alpha_c^{\pm}(L)$ in real-space ED up to $L=32$, we perform a linear extrapolation 
\begin{equation}
    \alpha_c^{\pm}(L) = \alpha_c^{\pm} + \frac{\beta}{L}
\end{equation}
to the thermodynamic limit in the inset of \figref{fig_dsEG}, obtaining critical breakdown values of $\alpha_c^- = -(1.556 \pm 0.018)$ and $\alpha_c^+ = 0.596 \pm 0.011$.

These critical points reveal a pronounced asymmetry in spinon stability for $J_1>0$ versus $J_1<0$, mirroring the field-theoretic predictions. Furthermore, this breakdown demonstrates that the entanglement transition, defined here by the scrambling of non-local fractionalized properties, is fundamentally distinct from the traditional thermodynamic phase boundaries for the spiral and dimer phases. As established in the broader context of quantum many-body systems~\cite{Thomale2_2010,Lundgren_2014}, entanglement spectra can exhibit abrupt topological reorganizations even when bulk thermodynamic observables show no signatures of a phase transition.  Our results establish that the definitive boundary of spinon confinement is dictated by a sharp macroscopic shift in the ground-state momentum character, which instantly drives a complete reorganization of the system's non-local entanglement structure independent of local thermodynamic order. Particularly in the ferromagnetic regime ($J_1 < 0$), where recent studies have characterized the formation of multi-magnon bound states~\cite{nishimoto_2025_systematic,agrapidis_2025_selforganized}, we propose that the spinons confine into these multi-magnons at the identified momentum transitions. Physically, this confinement mechanism would drive a transfer of spectral weight in the spectral function $S(q,\omega)$, abruptly transitioning from a broad, multi-spinon continuum into rather sharp, dispersive bound states forming a magnon.

\section{Conclusion and discussion}
\label{sec:conclusion}

In this work, we have demonstrated that the deconfined spinon description of the $J_1$--$J_2$ zig-zag spin chain exhibits an asymmetric robustness away from the exactly solvable limits. Through an analysis of the ES, we established that in the single-chain limit $J_1 \gg \abs{J_2}$, the maximum EG serves as a precise diagnostic for the system's proximity to the Haldane--Shastry point. In this regime, spinons survive across the liquid-dimer transition until their eventual breakdown at the Majumdar-Ghosh point.

Extending this to the double-chain zig-zag regime, we found that the ground state is accurately described by a double-spinon framework over a broad parameter space. By identifying the entanglement transition with a sharp shift in the ground-state momentum character, we find a tentative proxy for the boundaries of spinon confinement. We observe that the closure of the EG is dictated by an inter-chain excitation continuum, rather than intra-chain level renormalization. This transition thus potentially hints at a fundamental decoupling between the topological properties encoded in the ES including its effective quasiparticles and the system's thermodynamic phase boundaries.

The pronounced asymmetry in spinon stability---where ferromagnetic coupling ($J_1 < 0$) sustains fractionalized excitations into the spiral phase, while antiferromagnetic interchain coupling ($J_1 > 0$) confines spinons inside the dimer phase---provides a stringent test for current theoretical models. While coupled $\mathrm{SU}(2)_1$ WZW frameworks~\cite{White_1996,allen_1997_nonabelian} provide a beautiful qualitative explanation for this asymmetry, our quantitative entanglement results challenge a strict interpretation of these continuum field theories. By demonstrating that the spinon description remains valid well inside the QLRO($\pi/2$) phase, we show that the breakdown of fractionalization does not simply coincide with the QFT phase boundary located already at $J_1=0$.

Our results may carry significant implications for a variety of experimental platforms. In the $J_1 < 0$ regime, edge-sharing cuprates such as $\text{LiCuVO}_4$~\cite{enderle_2005_quantum} and $\text{LiCuSbO}_4$~\cite{dutton_2012_quantum,grafe_2017_signatures} serve as primary candidates for observing the extended spinon stability we predict within the spiral phase. Meanwhile, the $J_1 > 0$ confinement transition can be systematically explored in organic molecular magnets, such as the radical chain $\text{F}_2\text{PNNNO}$~\cite{hosokoshi_1999_magnetic} or more recently synthesized hexagonal-plaquette systems~\cite{yamaguchi_2025_realization}. Unlike rigid transition-metal oxides, the magnetic exchange in these organic radicals is mediated by noncovalent interactions within a soft solid-state lattice~\cite{paul_2021_magnetic}. This extreme structural sensitivity allows external stimuli, such as applied pressure or temperature variations, to reversibly alter distances between radical sites, enabling experiments to dynamically tune the $J_1/J_2$ ratio in situ. Consequently, these organic systems serve as the premier platforms for systematically sweeping across the sharp momentum-character shifts we identify as the signature of spinon breakdown.

Looking forward, our results open a promising avenue for deriving rigorous field-theoretic descriptions of the elusive spiral phase. Because the double-spinon picture remains valid well into this regime, future analytical work could construct the spiral ground state directly from these low-energy, fractionalized degrees of freedom. Furthermore, exploring the physical dynamical signatures of this delayed confinement will be crucial. While the inter-chain continuum we identify strictly governs the entanglement Hamiltonian $H_E$~\cite{Li_2008}, we expect this entanglement transition to drive a profound restructuring of the physical excitation spectrum $S(q, \omega)$.

Future numerical calculations of the dynamical structure factor, alongside high-resolution inelastic neutron scattering~\cite{brinckmann_2001_renormalized,bourges_2005_resonant,bera_2017_spinon,wang_2018_experimental,bera_2020_dispersions,ma_2022_magnetic} or 2D coherent spectroscopy~\cite{novelli_2020_persistent,hart_2023_extracting,fiore_2025_twodimensional} on the aforementioned materials, could track the transfer of spectral weight from a broad, multi-spinon continuum into sharp, confined magnon bound states~\cite{nishimoto_2025_systematic,agrapidis_2025_selforganized}. This bridge between theoretical entanglement measures and physical dynamics will provide a definitive, measurable footprint of the spinon breakdown we observe, potentially revealing the ``hidden'' entanglement transitions for which traditional thermodynamic probes might not be sensitive.

\section*{Acknowledgements}
We dedicate this work to the memory of Ian Affleck. We are grateful to Nicolas Regnault for providing insights into exact diagonalization methods within the spinon basis. We also thank Chris Hooley for stimulating discussions and for sharing insights regarding his work~\cite{mcroberts_2025_transition,azad_2025_generalized}.
\paragraph{Funding information}
This work is supported by the Deutsche Forschungsgemeinschaft (DFG, German Research Foundation) through Project-ID 258499086 -- SFB 1170 and through the W\"urzburg-Dresden Cluster of Excellence on Complexity and Topology in Quantum Matter -- ctd.qmat Project-ID 390858490 -- EXC 2147. T.He. was supported by the Deutsche Forschungsgemeinschaft (DFG, German Research Foundation) under Project No. 537357978 and, in part, by the US Department of Energy, Office of Basic Energy Sciences, Division of Materials Sciences and Engineering, under Contract No. DE-AC02-76SF00515. The Flatiron Institute is a division of the Simons Foundation.

\newpage

\begin{appendix}
\numberwithin{equation}{section}

\section{Bosonic mapping and momentum-space representation}\label{appendix: bosonic mapping and momentum-sapce representation
}
\subsection{Transformed Hamiltonian in the spinon description}\label{appendix: transformed hamiltonian in the single-spinon formalism}
We reformulate the $J_1$--$J_2$ Heisenberg Hamiltonian (see \eqnref{main_eq: real-space Hamiltonian} main text) in terms of canonical bosons with an additional onsite interaction to enforce a hardcore constraint. The transformed Hamiltonian reads
\begin{equation}
    H = \sum_{n=1}^2 J_n \sum_{i=1}^L \left[ \frac{1}{2} \left( \bdagger_i \bop_{i+n} + \text{H.c.} \right) + \left(\nop_i - \frac{1}{2}\right) \left(\nop_{i+n} - \frac{1}{2}\right) + V\left(\nop_i - \frac{1}{2}\right)^2 \right]\,,
\end{equation}
where $\nop_i = \bdagger_i \bop_i$ and $V$ is the added hardcore potential to prevent double occupancy in the limit $V\to \infty$. In our numerical analysis, we chose $V = 10^3$. Using the discrete Fourier transform of the bosonic operators defined in the main text, the Hamiltonian can be decomposed into a quadratic kinetic sector and a quartic scattering term. Up to constant energy shifts, the momentum-space representation is given by
\begin{subequations}
\begin{equation}
    H = \sum_{k} \varepsilon(k) \nop_k + \frac{1}{L} \sum_{k_1,k_2,k_3,k_4} \Gamma(\Delta k) \bdagger_{k_1} \bop_{k_2} \bdagger_{k_3} \bop_{k_4} \delta_{k_1+k_3, k_2+k_4}\,,
\end{equation}
where $\Delta k = k_4-k_3$ is the momentum transfer. The single-particle dispersion $\varepsilon(k)$ is defined~as
\begin{align}
    \varepsilon(k) &= \sum_{n=1}^2 \left[ J_n \left(\cos(nk) - 1\right) - V \right]\,,
\intertext{while the interaction vertex $\Gamma(\Delta k)$ accounts for density-density correlations and takes the form}
    \Gamma(\Delta k) &= \sum_{n=1}^2 \left[ J_n \cos(n \Delta k) + V \right]\,.
\end{align}
\end{subequations}
Note that the scattering term is subject to the constraint of crystal momentum conservation, i.e., the delta function $\delta_{k_1+k_3, k_2+k_4}$ is understood to be defined modulo $2\pi$.

\subsection{Transformed Hamiltonian in the double-spinon description} \label{appendix: transformed hamiltonian in the double-spinon formalism} 
In the regime of weak NN coupling $|J_1|\ll J_2$, it is convenient to treat the zig-zag chain as two coupled Heisenberg antiferromagnets corresponding to the even and odd sublattice, respectively. In this framework, we introduce two flavors of bosonic operators, $a_i$ and $b_i$, acting on the two subchains. The transformed bosonic Hamiltonian can be written as
\begin{align}
    H = J_1 &\sum_{i=1}^{L/2} \left[\frac{1}{2} \left( \adagger_{i} \bop_i + \adagger_{i+1} \bop_i + \text{H.c.}\right) + \left(n_{i}^a- \frac{1}{2}\right) \left(n_i^b - \frac{1}{2} \right) + \left(n_{i+1}^a- \frac{1}{2}\right) \left(n_i^b - \frac{1}{2} \right)\right] \nonumber \\
    +J_2 &\sum_{i=1}^{L/2} \left[ \frac{1}{2} \left( \adagger_i \aop_{i+1} + \bdagger_i \bop_{i+1} + \text{H.c.} \right) + \left(n_i^a - \frac{1}{2}\right) \left(n^a_{i+1} - \frac{1}{2}\right) + \left(n_i^b - \frac{1}{2}\right) \left(n^b_{i+1} - \frac{1}{2}\right) \right] \nonumber \\
    +V &\sum_{i=1}^{L/2} \left[ \left(n^a_{i} - \frac{1}{2}\right)^2 + \left(n^b_{i} - \frac{1}{2}\right)^2 \right]\,,
\end{align}
where $n^a_i = \adagger_i \aop_i$ and $n^b_i = \bdagger_i \bop_i$ denote the occupation numbers on the two subchains.
To construct a momentum-space representation of the Hamiltonian, we insert the discrete Fourier transform of both bosonic flavors (see \eqnref{main_eq: fourier transform of bosonic operators} main text). 

After dropping an overall constant, we obtain
\begin{subequations}
\begin{align}
H = \sum_{k} \bm{\Phi}_k^\dagger \mathbf{M}(k) \bm{\Phi}^{\phantom{\dagger}}_k + \frac{1}{L} \sum_{k_1, k_2, k_3, k_4} \sum_{\alpha, \beta} \Gamma^{\phantom{*}}_{\alpha\beta}(\Delta k) \sigmadagger_{k_1,\alpha} \sigmaop_{k_2,\alpha} \sigmadagger_{k_3,\beta} \sigmaop_{k_4,\beta} \delta_{k_1+k_3, k_2+k_4}\,,
\end{align}
where $\bm{\Phi}_k^{\phantom{\dagger}} = (\aop_k, \bop_k)^{\mathrm{T}}$ and $\sigma_{k,\alpha}$ represents the bosonic operators with flavors $\alpha=\{a,b\}$. The single-particle dynamics is encoded in the Hermitian kernel $\mathbf{M}(k)$, with matrix elements
\begin{align}
M^{\phantom{*}}_{\alpha\alpha}(k) &=  J_2 \left(\cos(k) - 1\right) - J_1 - V\,, \qquad
M^{\phantom{*}}_{ab}(k) = M_{ba}^*(k) = \frac{J_1}{2} (1 + e^{ik})\,, \\
\intertext{while the vertex $\boldsymbol{\Gamma}(\Delta k)$ describes the inter- and intra-flavor density-density correlations via}
\Gamma^{\phantom{*}}_{\alpha \alpha}(\Delta k) &= J_2 \cos(\Delta k) + V\,, \qquad \qquad \quad \,
\Gamma^{\phantom{*}}_{ab}(\Delta k) = \Gamma_{ba}^*(\Delta k)= \frac{J_1}{2} \left(1 + e^{i\Delta k}\right)\,.
\end{align}
\end{subequations}

\section{Measures of spinon deconfinement}\label{appendix: measures of spinon confinement}
\begin{figure}[h!]
    \centering
    \includegraphics[width=\linewidth]{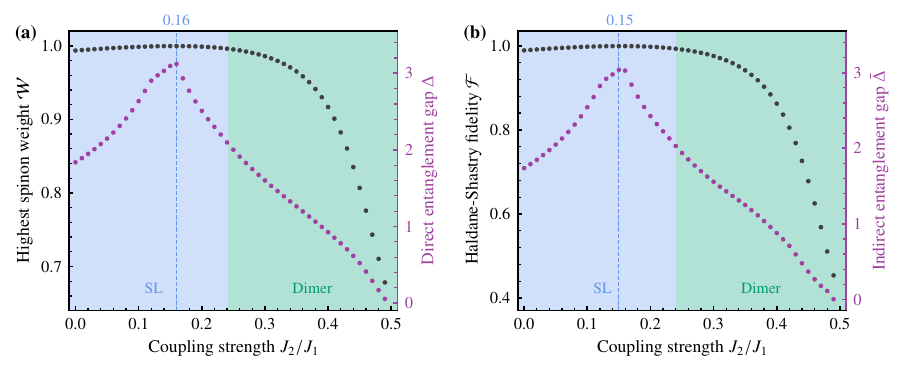}
    \caption{Measures of spinon deconfinement for $L = 22$, with a momentum cut containing $N_A = 6$ particles. (a) Highest spinon weight $\mathcal{W}$ and direct entanglement gap $\Delta$ as a function of $J_2/J_1$. The vertical blue dashed line marks the maximum of $\Delta$ and~$\sweight$ at $0.16$. (b) Haldane--Shastry fidelity $\mathcal{F}$ and indirect entanglement gap $\bar{\Delta}$, with the maximum of $\bar{\Delta}$ and~$\mathcal{F}$ being shifted slightly to $0.15$. Across both definitions, the observables are maximized at finite $J_2$, marking the regime where the Heisenberg antiferromagnet most closely resembles the Haldane--Shastry chain.}    
    \label{fig:spinon_measurement}
\end{figure}

To assess the similarity between the ground state $\ket{\Psi_0}$ of the $J_1$--$J_2$ Heisenberg chain and that of the Haldane--Shastry model,~$|\Psi_0^\mathrm{HS}\rangle$, we evaluate their ground-state fidelity $\mathcal{F} = \left| \langle\Psi_0 | \Psi_0^{\mathrm{HS}} \rangle\right|^2$. This quantity, alongside the direct and indirect entanglement gaps (see \eqnref{main_eq: direct EG} and \eqnref{main_eq: indirect EG} of the main text) and the spinon weight (see \eqnref{main_eq: spinon weight} of the main text), is presented in \figref{fig:spinon_measurement} for $L=22$ and $N_A=6$ as a function of the frustration ratio $J_2/J_1$ across both the spin-liquid and dimerized regimes. Within our numerical resolution ($0.01$), we find that the direct gap $\Delta$ is maximized at the same coupling as the spinon weight $\sweight$, whereas the indirect gap $\Bar{\Delta}$ attains its maximum at the coupling where the Haldane--Shastry fidelity $\mathcal{F}$ is highest. Although these two extrema do not coincide exactly, their close proximity indicates that these distinct observables provide consistent and robust measures of spinon deconfinement.

\newpage
\section{Spectral flow within the double-spinon framework}\label{appendix: evolution of entanglement spectra} 
\begin{figure}[h!]
    \centering
    \includegraphics[width=1\linewidth]{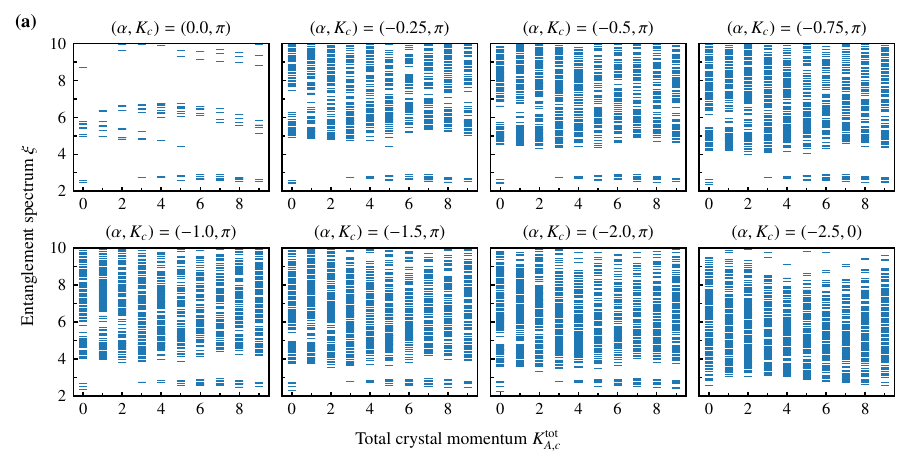}
    
    \vspace{0.5cm}
    
    \includegraphics[width=1\linewidth]{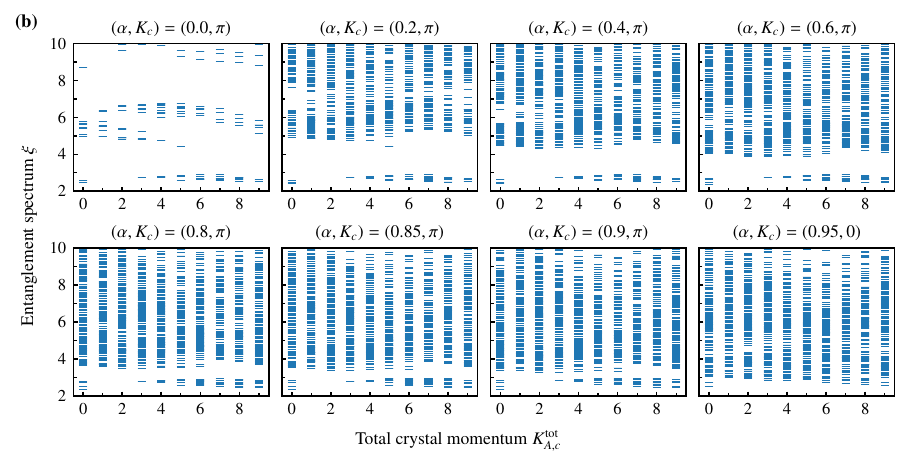}
    
    \caption{Entanglement spectra for $L_c=10$ unit cells, with a momentum cut containing $N_A=5$ particles. For representative values of the frustration coupling $\alpha=J_1/J_2$, the evolution from the decoupled limit to the spiral (a) and dimer (b) phase is shown. The double-spinon framework is protected by a finite entanglement gap, which collapses as the ground state momentum character shifts from $K_c=\pi$ to $0$. For the considered system size, this sharp transition occurs at $\alpha_c^-=-2.1403(4)$ and $\alpha_c^+=0.9261(1)$, respectively.}
    \label{fig:ent_spec_evolution}
\end{figure}

To illustrate the spectral flow from the decoupled limit ($J_1=0$) through the QLRO$(\pi/2)$ phase into the spiral ($J_1<0$) and dimerized ($J_1>0$) phases, we present a sequence of momentum-space entanglement spectra in \figref{fig:ent_spec_evolution}. The spectra partition into an $\mathrm{SU}(2)_1\times \mathrm{SU}(2)_1$ low-entangle\-ment part and non-universal higher levels, separated by a finite entanglement gap which prevents spinon confinement. For both signs of $J_1$, the double-spinon framework breaks down as the ground state momentum character shifts between $K_c=\pi$ and $0$.

\end{appendix}

\bibliography{refs.bib}

\end{document}